# High order synaptic learning in neuro-mimicking resistive memories

T. Ahmed,[1,][*] S. Walia,[1] E. L. H. Mayes,[2] R. Ramanathan,[3] V. Bansal,[3] M. Bhaskaran,[1] S. Sriram[1,][*] and O. Kavehei[1, †,][*]

[1]Functional Materials and Microsystems Research Group and Micro Nano Research Facility, RMIT University, Melbourne, VIC 3001, Australia

[2]RMIT Microscopy and Microanalysis Facility, RMIT University, Melbourne, VIC 3001, Australia

[3]Ian Potter NanoBioSensing Facility, NanoBiotechnology Research Laboratory, School of Science, RMIT University, Melbourne, VIC 3001, Australia

[†]Faculty of Engineering and Information Technologies, The University of Sydney, NWS 2006, Australia

* Corresponding authors. E-mail: taimurahmad1@gmail.com, sharath.sriram@gmail.com (S.S.), omid.kavehei@sydney.edu.au (O.K.)

## Abstract

Memristors have demonstrated immense potential as building blocks in future adaptive neuromorphic architectures. Recently, there has been focus on emulating specific synaptic functions of the mammalian nervous system by either tailoring the functional oxides or engineering the external programming hardware. However, high device-to-device variability in memristors induced by the electroforming process and complicated programming hardware are among the key challenges that hinder achieving biomimetic neuromorphic networks. Here, an electroforming-free and complementary metal oxide semiconductor (CMOS)-compatible memristor based on oxygen-deficient SrTiO$_{3-x}$ (STO$_x$) is reported to imitate synaptic learning rules. Through spectroscopic and cross-sectional transmission electron microscopic analyses, electroforming-free characteristics are attributed to the bandgap reduction of STO$_x$ by the formation of oxygen vacancies. The potential of such memristors to behave as artificial synapses is demonstrated by successfully implementing high order time- and rate-dependent synaptic learning rules. Also, a simple hybrid CMOS-memristor approach is presented to implement a variety of synaptic learning rules. Results are benchmarked against biological measurements form hippocampal and visual cortices with good agreement. This demonstration is a step towards the realization of large scale adaptive neuromorphic computation and networks.

## Introduction

The functionality of a brain is attributed to the activity-dependent synaptic weight change, enabling principal cognitive functions.[1] Although the underlying precise biological mechanism of the synaptic functionality is still under debate,[2] it is well established that *in vivo* neurons follow the Hebbian synaptic learning through spike-time-dependent-plasticity (STDP).[3-7] In order to mimic the biological synaptic learning, conventional complementary metal oxide semiconductor (CMOS) circuits have been employed,[8-9] but with no intrinsic learning capability, high energy consumption and limited scalability. All these factors prevent CMOS neuromorphic systems from even reaching towards the efficiency[10] and density (~$10^{11}$ neurons and ~$10^{15}$ synapses compact in volume of ~1130-1260 $cm^3$)[11] of the human brain.

Nanoscale memristors have attracted attention as potential artificial synapses due to their similar activity-dependent nonlinear conductance modulation.[12-15] However, memristors require integration with the driving CMOS subsystems to successfully execute the memory/computation operations and emulate synaptic functions. To date, several hybrid CMOS-memristor architectures have been reported to achieve high density memory systems and neuromorphic computing paradigm.[16-18] But complex CMOS circuitry, inexorable electroforming process causing a high device-to-device variability and stochastic nature of resistive switching are hampering the realization of efficient neuromorphic networks.[15-16, 19] A hybrid architecture implementing a simple dynamic CMOS circuitry to comply with any type of memristors and an energy efficient element would enable the imitation of versatile neuromorphic functions.

In this study, we exploit electroforming-free characteristics of $STO_x$ memristors and flexible design of a CMOS drive circuit to demonstrate the realization of an efficient neuromorphic system. Over the recent few years, a variety of synaptic functions including short- and long-term memory, paired-pulse facilitation and pair-based STDP ($p$-STDP) have been implemented on different types of memristors.[11, 20-22] However, higher order time-dependent plasticity such as triple-STDP ($t$-STDP) and quadruplet-STDP ($q$-STDP), and rate-dependent learning rules such as and Bienenstosk-Cooper-Munro (BCM)[23-25] synaptic modification (well known to exist in biological synapses) have not been experimentally demonstrated in artificial synaptic devices. Though the classical $p$-STDP models helped to establish a fundamental understanding of the Hebbian synaptic plasticity in several neural systems but it is not sufficient to accurately model all biological experimental results produced by multiple (triplet and quadruplet) spikes.[23, 26] This can be associated with deficiencies in the classical $p$-STDP model, such as excluding non-linear integration of spike pairs and their repetition frequency to quantify the synaptic modification.[24, 26] This infers the classical $p$-STDP model cannot elicit the BCM synaptic learning rule, which is regarded as a possible explanation of experience-dependent synaptic plasticity.[24] On the other hand, the $t$-STDP model is believed to be comprehensive enough to explain the biological experimental results produced by multiple spikes. As such, the implementation of the $t$-STDP rule on $STO_x$ synaptic devices can highlight the capability of these artificial synapses to mimic the biological synaptic functionalities. We acknowledge that a few memristive models and circuits have recently been proposed to reproduce these synaptic learning rules.[27-29] However, an experimental emulation of these essential biological learning rules will signify the potential of memristors for future neuromorphic computation.

## Switching Characteristics of STO$_x$ Synaptic Devices

The STO$_x$ synaptic devices are fabricated in a metal–insulator–metal (MIM) configuration as a bilayer stack of Ti/STO$_x$ sandwiched between top and bottom Pt electrodes. The stoichiometric analysis of STO$_x$ thin films shows that the sputtered films are oxygen-deficient which indicates presences of oxygen vacancies in the as-grown STO$_x$ switching layer (Supporting Information, Figure S1). **Figure 1**a shows an electroforming-free clockwise bipolar switching behavior of the STO$_x$ synaptic devices. As depicted in Figure 1a, a negative triangular DC voltage sweep with amplitude of -1 V (as $V_{SET}$) switches the MIM device from its high resistance state (HRS) to a low resistance state (LRS). While an opposite polarity triangular DC voltage sweep with an amplitude of +1.3 V (as $V_{RESET}$) switches the device to its HRS. This behavior *i.e.*, RESET on positive bias and SET on negative bias is typical for STO-based memristors.[30-33] The as-fabricated MIM devices are in their high resistive state (HRS) as the measured pristine resistances are close to the normal variance of HRS achieved during subsequent cyclic switching (Supporting Information, Figure S2). However, the pristine state resistances are cell area dependent. Furthermore, the statistical analysis of the as-fabricated devices shows that the average SET and RESET voltages during the first *I–V* sweeps are also area dependent (Supporting Information, Figure S2). The electroforming-free characteristic of the STO$_x$ synaptic devices can be associated with the presence of as-grown and increase in the concentration of oxygen vacancies under applied bias.[34-35] It is possible that during the first SET sweep Joule heating may induce additional oxygen vacancies in the MIM structure, according to oxygen exchange reaction ($O_O \rightleftharpoons \frac{1}{2}O_2 + V_O + 2e$).[36] As such, the increasing concentration of oxygen vacancies reduces their migration distance and consequently the electrical energy required to form a conductive path. Further evidence and cross-sectional characterisation of the filamentary path is given in the following sections.

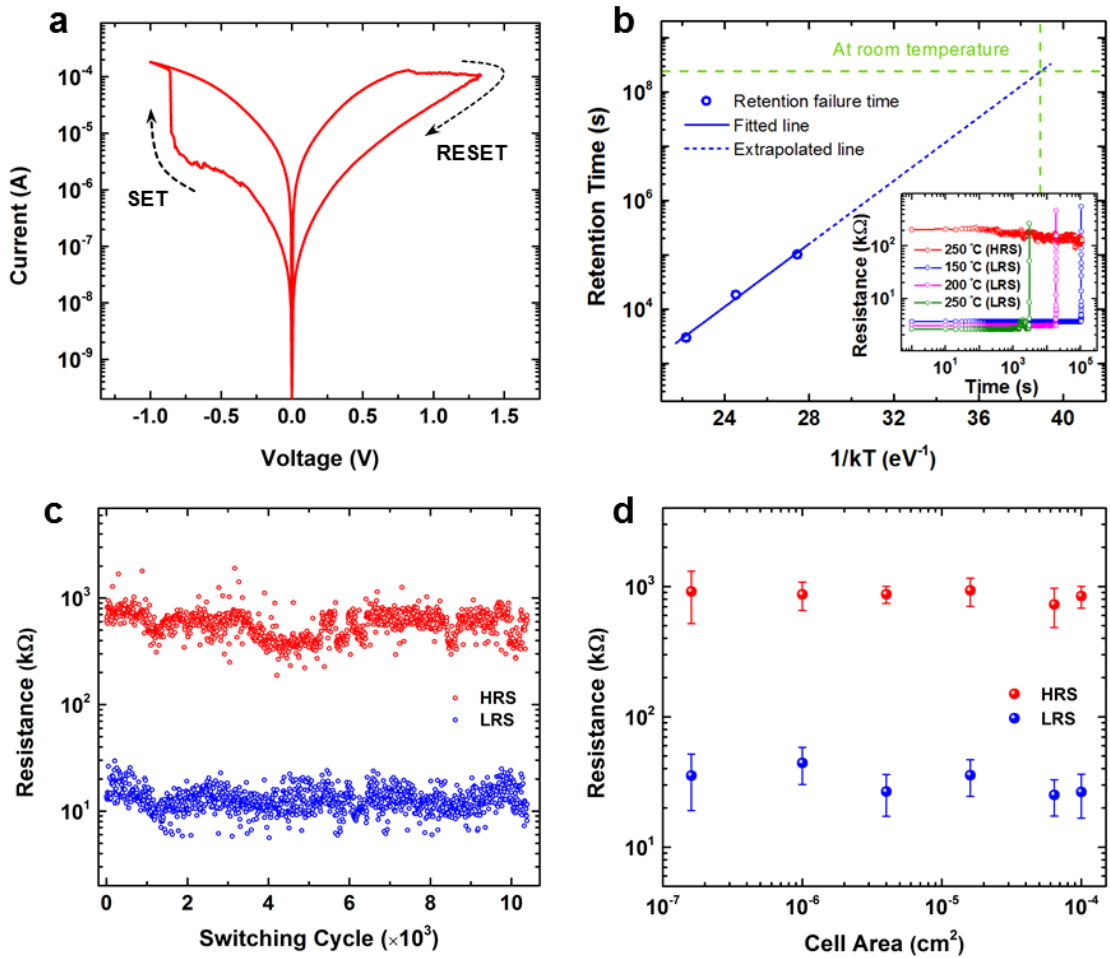

**Figure 1. Electrical characterisation of the STO$_x$ synaptic devices.** (a) The *I–V* characteristic sweep of a 10×10 μm$^2$ STO$_x$ MIM device. (b) The retention time *vs.* 1/kT plot to evaluate the state stability of the STO$_x$ devices. The inset shows retention of LRS and HRS at different elevated temperatures ranging from 150 to 250 °C. (c) Endurance of the devices, where $V_{RESET}$ of -1.6 V, $V_{SET}$ of +1.4 V and $V_{READ}$ of +0.1 V are applied as a train of short pulses. (d) The dependence of HRS and LRS on the active cell area.

To evaluate the reliability of the STO$_x$ MIM devices, the resistive states are measured at elevated temperatures ranging from 150 to 250 °C, as shown in the inset of Figure 1b. The retention of HRS measured for 30 hours at 250 °C shows no failure, indicating high stability of HRS. However, retention characteristics of LRS are temperature-sensitive. This high temperature LRS retention failure can be associated with the thermally-assisted reduction in the concentration of oxygen vacancies in the nano-filament and eventually its rupture.[37] The LRS retention failure time at different temperatures (where resistance jumps higher than the HRS) is plotted in an Arrhenius plot, as shown in Figure 1b, to calculate the oxygen vacancy migration activation energy and estimate the retention characteristics of the STO$_x$ memristors. The extrapolation of the fitting line in Figure 1b estimates the retention time of *ca.* 7.6 years at room temperature. Even though, this retention is suitable for memory and neuromorphic applications, it may be further improved by preventing the re-oxidation of STO$_x$ oxide layer through inserting a thin film exhibiting slow oxygen diffusion coefficient, such as Al$_2$O$_3$.[38] On the other hand, an activation energy of *ca.* 0.29 eV is extracted from the linear fitting of the experimental data points in Figure 1b. This lower LRS activation energy, as compared to the other oxide systems[37, 39-40] (*e.g.*, 1.0-1.6 eV reported for *a*-Al$_2$O$_3$, *a*-Ta$_2$O$_5$, *a*-Nb$_2$O$_5$, and TiO$_2$), suggests an hopping conduction mechanism in our STO$_x$ MIM devices.[41] This hopping conduction refers to the electronic transport through localized states, where these states are provided by the oxygen vacancies in the nano-filament.

To evaluate the switching repeatability of the $STO_x$ MIM devices (Figure 1c), short pulses of 1 μs duration and amplitude of -1.4 V and +1.6 V are applied for SET and RESET operations, respectively. READ pulses with amplitude of +0.1 V and duration of 200 ns are used to measure the SET/RESET currents. The effect of pulse width on the switching performance is also evaluated (Supporting Information, Figure S4). The endurance characteristics for more than $10^4$ switching cycles (Figure 1c) indicate that the synaptic devices exhibit repeatable bipolar switching behaviour. Typically in transition metal oxides, the bipolar resistive switching behavior is attributed to the inhomogeneous conduction mechanisms through the localized filamentary pathways and associated redox processes.[36, 42-44] As such, the resistance states (*i.e.*, HRS and LRS) are expected to be independent of the lateral dimensions of the MIM devices. Figure 1d reveals no appreciable area-dependency in our $STO_x$ devices for either resistance state. This further supports our earlier statement regarding the formation of conductive filamentary pathway in the MIM devices. In order to explain the electroforming-free switching behaviour of our $STO_x$ synaptic devices, we have conducted detailed material characterisation and cross-sectional microstructural analyses, as described in the following sections.

**Visualising Filamentary Switching in $STO_x$ Synaptic Devices**

The physical structure of the $STO_x$ synaptic devices and their compositional analysis is characterised by cross-sectional transmission electron microscopy (TEM). Electron energy loss spectroscopy (EELS) is used to assess the distribution of oxygen content in the MIM devices. The cross-sectional micrograph and corresponding EELS spectra of a pristine MIM device reveal an amorphous microstructure of the $STO_x$ layer and a partial oxidation of the top Ti layer at the Ti/STO$_x$ interface (Supporting Information, Figure S5). The amorphous and oxygen-deficient structure of STO$_x$ layer is attributed to the room temperature sputtering in a pure argon environment. Also, the partial oxidation of the top Ti layer to sub-oxide at Ti/STO$_x$ interface can be associated with the interfacial oxygen diffusion and Ti–O bonding between Ti and STO$_x$ oxygen ions.[45-47] **Figure 2**a,b shows scanning TEM (STEM) images of the switching STO$_x$ memristive devices in their LRS and HRS, respectively. High contrast regions are observed in the STO$_x$ layers and along the top Ti/STO$_x$ interfaces which indicate the applied electric field induced compositional changes in the STO$_x$ layers. To analyse the state-dependent composition of the STO$_x$ layers, region of interests (ROIs) are selected across the lamellae, highlighted in Figure 2a,b. The EELS O–*K* edge area maps (Figure 2c,d) show the relative distribution of oxygen content in ROIs where the area maps are generated by taking the O–*K* edge intensities of the collected spectra (at each pixel) after pre-edge background subtraction. The O–*K* edge area map of the device in LRS (Figure 2c) reveals the presence of an extending oxygen-deficient region between top and bottom Pt electrodes. This indicates a localised accumulation of oxygen vacancies and formation of conductive filamentary path across the MIM structure.[30] On the other hand, the O–*K* edge area map of the device in HRS (Figure 2d) shows higher concentration of oxygen vacancies at the vicinity of bottom Pt electrode, an indication of a ruptured filamentary path.

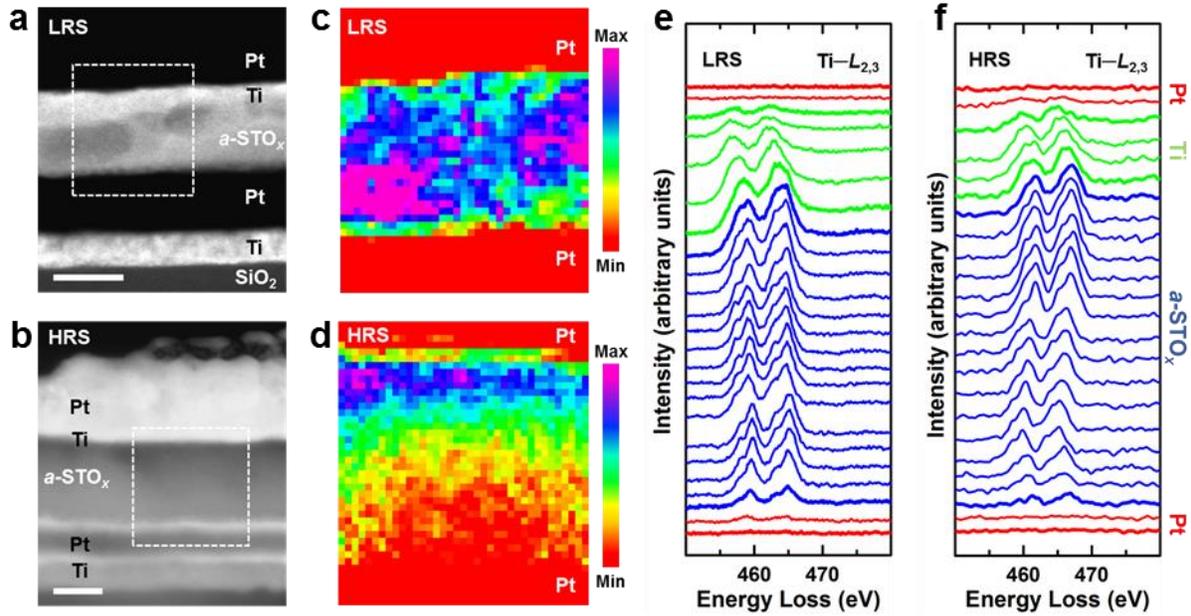

**Figure 2. Microstructural and compositional analyses of the STO$_x$ synaptic devices.** (a) STEM cross-section of a switching device in its LRS. Scale bar 20 nm. (b) STEM cross-section of a switching device in its HRS. Scale bar 20 nm. (c) The EELS O–$K$ edge area map of the enclosed region of interest in (a). The colour bar shows the relative oxygen content. (d) The EELS O–$K$ edge area map of the enclosed region of interest in (b). The colour bar shows the relative oxygen content. (e) The EELS Ti–$L_{2,3}$ edge profiles along a line scan across the ROI in (a). (f) The EELS Ti–$L_{2,3}$ edge profiles along a line scan across the ROI in (b).

The formation of each oxygen vacancy in STO introduces two electrons into the Ti $3d$ orbital, and the resulting change in the Ti valence can be observed in the EELS Ti–$L_{2,3}$ edge profile.[48-49] Figures 2e and 2f shows the background corrected Ti–$L_{2,3}$ spectra acquired along the EELS cross-sectional line scans passing over the ROIs indicated in Figures 2a and 2b, respectively. The Ti–$L_{2,3}$ fine structures gradually evolve in their intensity and position (from top Pt/Ti interface to the bottom Pt electrode) as clearly observed in both LRS and HRS. Due to resolution limitations, we evaluate the Ti–$L_{2,3}$ edge profiles to qualitatively analyse the electronic structure of the ROIs.

The broad and relatively low intensity peaks at top Pt/Ti interface indicate the presence of mixed $Ti^{2+}$ and $Ti^{3+}$ oxidation states which highlights the oxidation of Ti layer.[45, 47-48, 50] At Ti/STO$_x$ interfacial region and in the STO$_x$ layer, the crystal-field splitting of Ti–$L_3$ and Ti–$L_2$ peaks (into $t_{2g}$ and $e_g$ peaks) and their shift can be attributed to the presence of $Ti^{3+}$ and $Ti^{4+}$ oxidation states.[51-53] It is well known that, in transition metal oxide based resistive memories, the resistive switching is attributed to the redox reactions and associated valence change in the transition metal cations such as Ti in STO.[36, 54] As such, the cross-sectional TEM analyses show that the bipolar resistive switching in our STO$_x$ MIM devices is of filamentary nature where formation and rupture of extended oxygen-deficient regions and associated change in Ti valance result in LRS and HRS, respectively.

## Implementation of synaptic functions

A typical biological synapse consists of a pre-synaptic neuron and a post-synaptic neuron connected through a synapse, as schematically illustrated in **Figure 3a**. In a memristor based artificial synapse, the bottom and top electrodes work as neurons and the switching layer acts as a synaptic connection. The electrical conductivity of the device interprets the synaptic weight, while its increase or decrease translates to potentiation or depression, respectively, in response to the applied voltage spikes. Figure 3b shows an experimental implementation of simplified *t*-STDP learning rules as described in Ref.[23-24], using our STO$_x$ memristor.

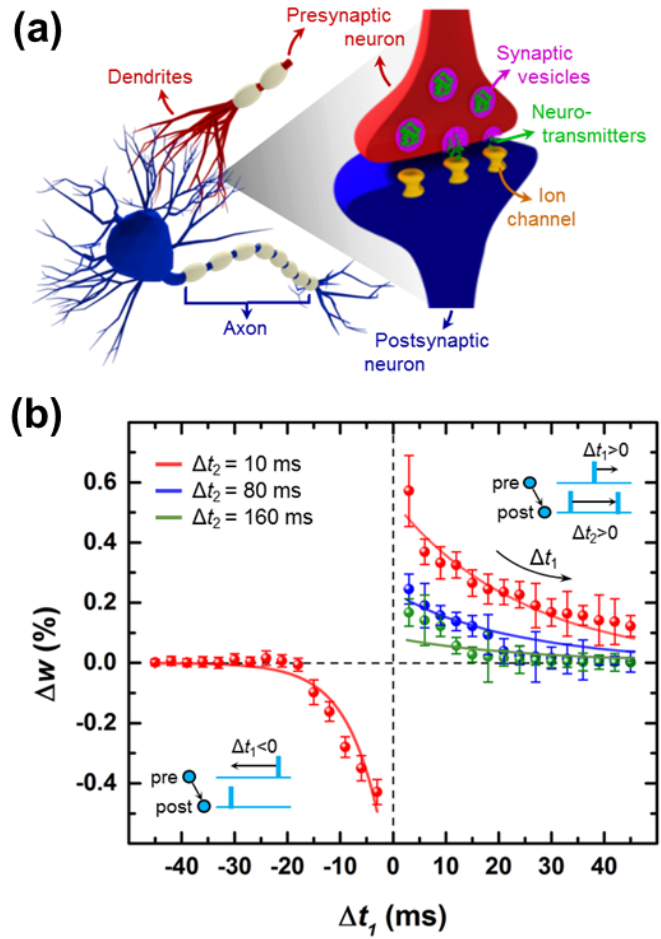

**Figure 3. Triplet-based STDP window implemented on STO$_x$ synaptic devices.** (a) An illustration of two biological neurons connecting *via* synapses. (b) Artificial implementation of STDP learning rules using STO$_x$ synaptic devices. Each data point and its deviation from mean (represented by bars) are collected by applying 100 cycles of identical pulses, where each cycle contains a RESET (for potentiation experiments) or SET (for depression experiments) pulse.

Synaptic changes reported here (Figure 3b) are collected using application of different pulse magnitudes with fixed pulse widths. A time-to-digital-to-voltage circuitry (shown in Figure 5a and discussed later) is simulated to generate magnitude of the voltage pulses corresponding to the spike-timing information ($\Delta t_1$ and $\Delta t_2$). To verify the capability of this scheme to implement a wide range of learning rules including *p*-STDP and *t*-STDP, we have applied a series of 100 pulses for each voltage magnitude that is chosen by the programming circuitry. The amplitude of applied voltage pulses for the corresponding spike-timing is listed in **Table S1,** Supporting Information. These experiments demonstrate a simple analog time-multiplexing implementation of artificial memristive synapses with shared peripheral circuitry.

A simplified *t*-STDP learning rule,[23-24] mentioned in Equation 1, suggests that synaptic depression is produced by spiking pairs with time interval of $\Delta t_1$ (as in classical *p*-STDP rule), while synaptic potentiation takes a triplet of spikes into account. Here, we consider post-pre-post configuration of the triplet spikes, all details can be similarly applied for a pre-post-pre configuration. It is also important to note that the asymmetry of the STDP window, shown in Figure 3b, is due to the asymmetry in potentiation and depression rates of the $STO_x$ memristors and is consistent with several *in vivo* and *in vitro* STDP studies conducted on different types of biological synapses.[5]

A simplified *t*-STDP learning rule can be shown as,[24]

$$\begin{aligned}
\Delta w^-(\Delta t_1) &= -A_1^- e^{(\Delta t_1/\tau_-)}, & \Delta t_1 < 0 \\
\Delta w^+(\Delta t_1, \Delta t_2) &= A_2^+ e^{(-\Delta t_1/\tau_+)} e^{(-\Delta t_2/\tau_y)}, & \Delta t_1 \geq 0, \Delta t_2 \geq 0
\end{aligned} \quad (1)$$

where $\Delta t_1 (= t^{post} - t^{pre})$ and $\Delta t_2 (= t^{post1} - t^{post2})$ are time differences. $A_1^-$ and $A_2^+$ are constant amplitudes of each exponential term in potentiation ($\Delta w^+$) and depression ($\Delta w^-$) equations. The values of these amplitudes extracted from curve fitting (in Figure 3b) are $A_1^- = -0.70$ and $A_2^+ = 0.60$. Also, $\tau_+$ and $\tau_-$ are time constants of $\Delta w^+$ and $\Delta w^-$, respectively, and obtained from the fitting parameters as $\tau_+ = 8.2$ ms and $\tau_- = 2.5$ ms. While the time constant $\tau_y$ indicates the exponential correlation between $\Delta w^+$ and $\Delta t_2$, and extracted as $\tau_y =$ 80 ms. To reproduce the *t*-STDP window, the values of $\Delta t_2$ are fixed at 10, 80 and 160 ms during the experiments (as shown in Figure 3).

In order to demonstrate that our STO$_x$ synaptic devices are capable to imitate the biological synaptic plasticity, we implement the *t*-STDP model (Equation 1) by following the experimental protocols reported by Pfister and Gerstner[23] and compare the results with the electrophysiological experiments performed in hippocampal culture[55] and visual cortex[25] (Figure 4). Two different triplet spiking patterns, namely post-pre-post (*i.e.*, 1-pre-2-post) and pre-post-pre (*i.e.*, 2-pre-1-post), are used in hippocampal culture experiments,[55] each consists of 60 triplet of spikes and repeated at a given rate (1 Hz). The weight change as a function of timing difference between pre- and post-synaptic spikes in both triple patterns is graphically presented in Figure 4a and Figure 4b. The best fit is calculated by a normalized mean-square error function (*E*) represented as,[23]

$$E = \frac{1}{P}\sum_{i=1}^{P}\left(\frac{\Delta w_i^{exp} - \Delta w_i^{mem}}{\sigma_i}\right)^2 \qquad (2)$$

where *P*, $\Delta W_i^{exp}$, $\Delta W_i^{mem}$ and $\sigma_i$ are the number of data points in a dataset, mean weight change (in electrophysiological and STO$_x$ memristor experiments) and the standard error mean (SEM) of $\Delta W_i^{exp}$ for a given data point *i*, respectively. In the hippocampal culture, 13 data points are used, which includes 2 pairing and 3 quadruplet data points. To compare our experimental results with hippocampal culture, we use only 8 triplet data points, 4 for 2-pre-1-post and 4 for 1-pre-2-post triplet spiking patterns. Also, we minimize the function *E* (given in Equation 2) which represents the error between our memristor experimental results and hippocampal culture (Figure 4a and 4b). Conventionally, parameters of the CMOS drive circuit are tuned to achieve the best match between the mathematical *t*-STDP and hippocampal culture data.[27-28] However, we minimize the error *via* one to one mapping of the weight changes ($\Delta w$) to the appropriate voltage levels applied to the artificial synapses ($STO_x$ memristors). Furthermore, this mapping is carried out by extracting timing information using a time-to-digital (T2D) and then digital-to-voltage conversion in the CMOS drive circuitry (as explained below). As such, the error is minimized by creating and changing a digital look-up table that maps incoming spike-timings information to a 6-bit digital code. The weight change corresponding to the both triplet pairing configurations is listed in **Table S2**, Supporting Information. There is <10% error between our memristor data and hippocampal culture data which is comparatively smaller than previously achieved by mathematical *t*-STDP model.[27-28]

Figure 4c shows the implementation of BCM learning rule where the synaptic weight changes as a function of the given frequency, $\rho$. The comparison of our experimental results with the visual cortex data set (Figure 4c) shows that $STO_x$ memristors closely follow the BCM behavior for $\rho \leq 30$ Hz, while for high frequencies our experimental results are within the variation limits of visual cortex data set. As observed in the *I-V* characteristics of the $STO_x$ memristors, SET process exhibits a digital-like behavior while RESET is an analog-like switching behavior offering comparatively more intermediate stable-states. As such, achieving a high dynamic range of weight change for $\Delta w^+$ is more challenging than $\Delta w^-$, also observed in Figure 3b and Figure 4c. The values of synaptic weight change corresponding to different frequencies are listed in **Table S3**, Supporting Informed. This indicates that like time-dependent learning rules (*i.e.*, *p*-STDP and *t*-STDP); the BCM rule can also be implemented by our $STO_x$ synaptic devices.
14

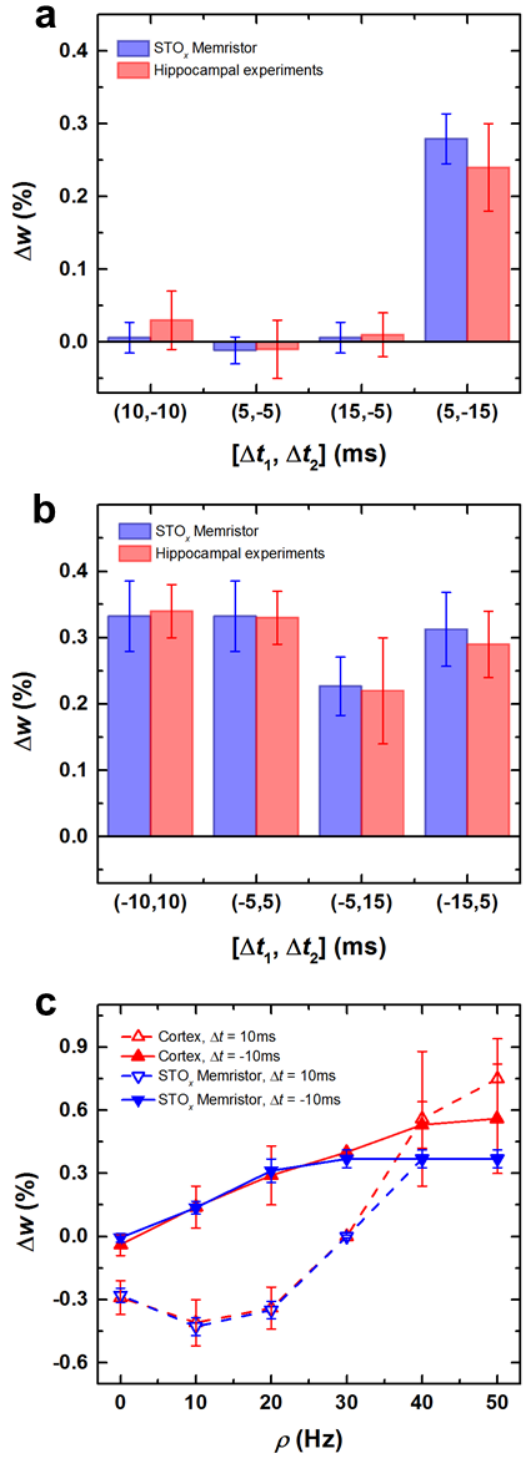

**Figure 4. Reproduction analyses of the time- and rate-dependent learning rules.** The reproduction of weight change induced by (a) pre- post- pre and (b) post- pre- post triplet spike patterns. (c) The synaptic weight change as a function of spike rate.

## CMOS Drive Circuitry

The available literature reports either extensive peripheral circuitry to generate suitable pre- and post-synaptic spike shapes (similar to the biological action potentials) or special circuits designed for a specific type of memristive system.[16, 56] Additionally, the direct interfacing between CMOS drive circuitry and memristive devices/array can expose them to CMOS circuit non-idealities.[57] Herein, we utilize a well-established CMOS circuit, called forward body biasing,[58-61] in combination with a time-to-digital converter to implement not only time dependent synaptic rules but also demonstrate the potential of implementing a wide variety of synaptic learning rules.

**Figure 5a** shows a schematic of the proposed CMOS drive circuit which is a modification of the body-bias generator,[59-60] and converts differences in input spike-timing into voltage amplitudes. The T2D module is responsible for the pre- and post-synaptic event digitization and includes a timing control unit and a decoder (see Section S5, Supporting Information). The timing control unit is a fully digital unit that receives pre- and post-spikes and generates a binary code according to the timing intervals and works based on a number of counters that are triggered and stopped with spikes. It can be configured for multiple protocol implementation and $\Delta t$ detection.[58] As depicted in Figure 5a, a finely tuned voltage ($V_w$) is generated to modify the weight of a memristor and is connected to a memristor array *via* a voltage follower (VF) and an array of transmission gates (TGs) that are connected to the top-electrodes (TEs) and bottom-electrodes (BEs) of the memristor array. It is worth mentioning that we focus on design of the peripheral circuitry for memristive artificial synapses, while modifications in the neuron designs may also be necessary to consider online-learning aspects of the learning rules discussed above. Each device in the array is individually accessible *via* an addressable top-electrode (TE) and bottom-electrode (BE) connections. Selections are mandated externally and also partially include some internal data. These selection signals are represented with two digital vectors for rows (R) and columns (C), *i.e.*, $Sel_R$ and $Sel_C$ respectively, in Figure 5a. Note that $I_{ref}$ is a constant reference current that is supplied through a digital to analog converter (DAC), while $V_{DDA}$ and $V_{SSA}$ represent analog voltage supply and ground of the drive circuit, respectively. The proposed drive circuitry disconnects timing scales from the voltage level generation. Also the T2D module is fully programmable and is capable of mapping any spike-timing to any binary code which can be translated to the corresponding voltage amplitude, *via* the DAC.

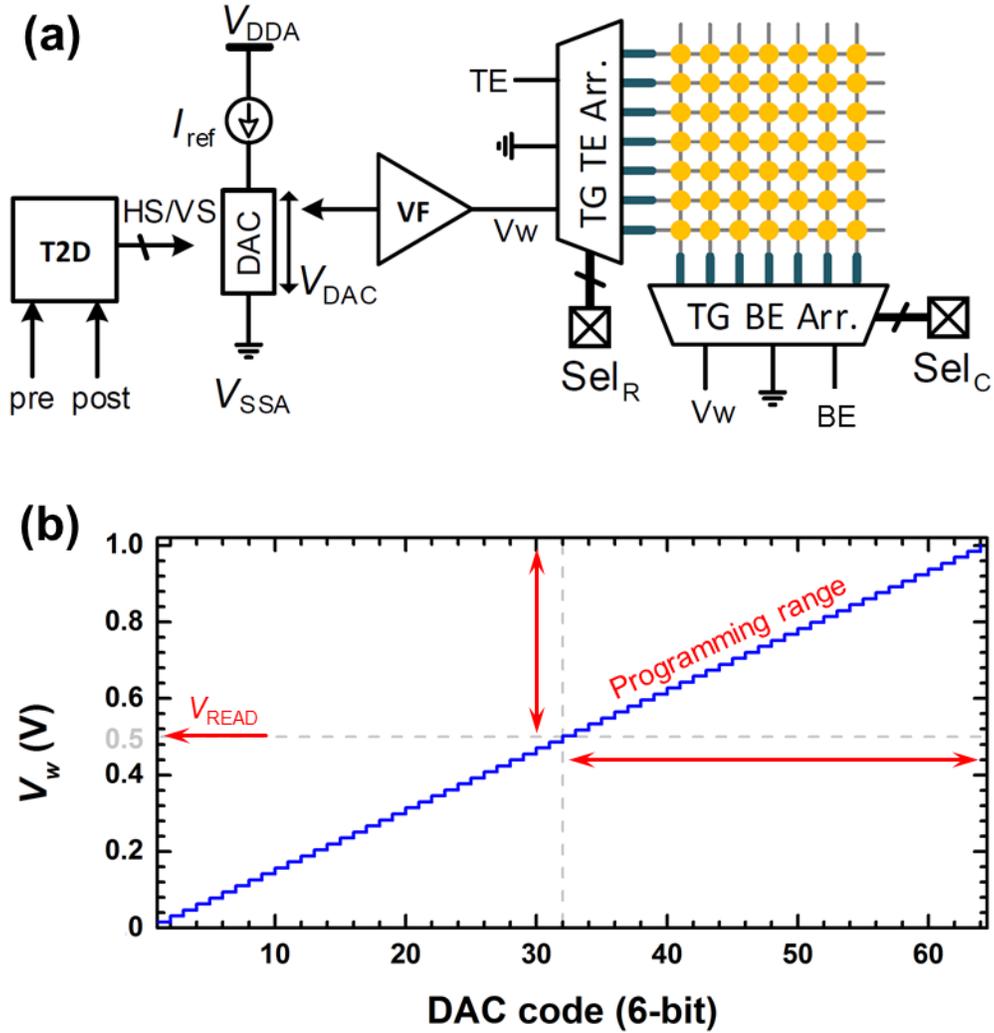

**Figure 5. The CMOS drive circuitry.** (a) A schematic of the proposed CMOS drive circuit which converts difference in input spike-timing into voltage amplitudes to modify the synaptic weight of a target memristor in the array. (b) Simulated resolution of the DAC circuitry to generate the weight changing voltage, *i.e.*, $V_w$.

As revealed in Figure 3b, near exponential relationship exists between the programming voltage amplitude ($V_w$) and synaptic weight change ($\Delta w$). This implies that a small variation in $V_w$ can cause a significant deviation in $\Delta w$. Therefore, it is essential to estimate the programming efficiency of the proposed CMOS drive circuit. Figure 5b shows the Cadence simulation of the DAC circuitry using 90 nm CMOS technology. A 15.6 mV resolution of the $V_{DAC}$ for a total 1 V supply is achieved. Figure 5b and Table S1 (Supporting Information), shows overall mapping of spike-timing to DAC code and then to an equivalent voltage ($V_w$ in this case). It has been reported that variation in $V_w$ is less than 5 mV.[59-60] Although the 15.6 mV increase in applied voltage magnitudes even higher than the $V_{READ}$ (*i.e.,* 0.5 V) do not necessarily switch the device, it is observed that such an increase makes a significant statistical change in $\Delta w$.

## Conclusion

In summary, we have presented a CMOS-compatible memristor based on $STO_x$ exhibiting electroforming-free bipolar resistive switching behaviour. First, through electrical and cross-sectional characterisations we have shown that reliable resistive switching in the $STO_x$ memristors is attributed to the redox reactions and electronic transport through the localised conductive filamentary pathway. In addition to the electroforming-free characteristics, the $STO_x$ memristors have also shown their scalability potential for future high-density memory applications. Secondly, we have demonstrated a hybrid CMOS-memristor approach to successfully mimic high order time-dependent, such as *p*- and *t*-STDP, and rate-dependent, such as BCM, synaptic learning rules. As such, this study is a step towards the realization of an adaptive neuromorphic network.

## Experimental

*Device fabrication:* The STO$_x$ synaptic devices are fabricated as cross-point devices and array in metal-insulator-metal (MIM) configuration with active areas of 2×2 μm$^2$, 4×4 μm$^2$, 10×10 μm$^2$, 20×20 μm$^2$, 40×40 μm$^2$, 80×80 μm$^2$ and 100×100 μm$^2$ by following standard photolithography and thin film deposition processes. The bottom Pt (15 nm)/Ti (7 nm) electrodes are patterned onto a SiO$_2$ (300 nm)/Si substrate by electron beam (e-beam) evaporation. As a switching layer, 25 nm thin film of amorphous oxygen-deficient STO is deposited by using radio frequency sputtering (with 100 W power) from a commercial STO ceramic target in a pure argon environment under a pressure of 0.46 Pa and at room temperature. In order to complete the MIM structure, top Pt (30 nm)/Ti (5 nm) electrodes are evaporated by the e-beam evaporation at a base pressure of <6×10$^{-5}$ Pa.

*Electrical characterisation:* The electrical characterisation of the STO$_x$ synaptic devices is performed under ambient conditions by using the Keithley 4200SCS equipped with remote preamplifiers and 4225 pulse modulation unit and Agilent 2912A sourcemeter connected to a micro-probe station. High temperature electrical measurements are performed by using an environmentally isolated Linkum chamber connected with Agilent 2912A sourcemeter.

*X-ray photoelectron spectroscopy:* The X-ray photoelectron spectroscopy (XPS) spectra are collected by using a Thermo Scientific K-Alpha instrument equipped with an Al *K*α radiation source (1486.6 eV). For the XPS analysis bare STO$_x$ oxide thin films (~25 nm) sputtered on SiO$_2$/Si substrates and a commercial stoichiometric STO (100) single crystal substrate (as a reference) are used. All spectra are resolved by using the standard Gaussian-Lorentzian function followed by the Shirley background correction.

*Photoluminescence spectroscopy:* The photoluminescence emission spectrum is obtained using Horiba Scientific FluoroMax-4 spectrofluorometer, at room temperature from. A laser source with 325 nm of wavelength is used to excite the sputtered STO thin film.

*Cross-sectional analyses:* The transmission electron microscopy (TEM) and electron energy loss spectroscopic (EELS) analyses are performed on pristine and switching $STO_x$ MIM devices (at least for 50 cycles and subjected to constant bias stresses) using a JEOL 2100F scanning transmission electron microscope (STEM) with attached Tridium Gatan image filter with an aperture of 5 mm. The TEM samples are prepared by focused ion beam cuts through the MIM structure by using a FEI Scios DualBeam$^{TM}$ system. Cross-sectional STEM micrographs and EELS spectra are collected using a <1.5 nm beam spot. EELS spectra are collected with a dispersion of 0.3 eV per pixel which allowed simultaneous recording of the titanium $L_{2,3}$ (Ti–$L_{2,3}$) edge and oxygen $K$ (O–$K$) edge in the regions of interest and across the MIM cross-sections. A power law fit is adopted for the pre-edge background correction while the influence of nearby peaks and plural scattering are reduced by narrow signal windows.

**Acknowledgements**

The authors acknowledge support from the Australian Research Council (ARC) for personnel and project support via DP130100062 (S.S.), DE160100023 (M.B.), and FT140101285 (V.B.) and equipment funding through LE0882246, LE0989615, LE110100223, and LE150100001. The authors would like to acknowledge the technical assistance of the Micro Nano Research Facility (MNRF) and the RMIT Microscopy and Microanalysis Research Facility (RMMF).

# Supporting Information

## High order synaptic learning in neuro-mimicking resistive memories

*T. Ahmed,[1,]\* S. Walia,[1] E. L. H. Mayes,[2] R. Ramanathan,[3] V. Bansal,[3] M. Bhaskaran,[1] S. Sriram[1,]\* and O. Kavehei[1, †,]\**

[1]Functional Materials and Microsystems Research Group and Micro Nano Research Facility, RMIT University, Melbourne, VIC 3001, Australia

[2]RMIT Microscopy and Microanalysis Facility, RMIT University, Melbourne, VIC 3001, Australia

[3]Ian Potter NanoBioSensing Facility, NanoBiotechnology Research Laboratory, School of Science, RMIT University, Melbourne, VIC 3001, Australia

†Faculty of Engineering and Information Technology, The University of Sydney, NWS 2006, Australia

\* Corresponding authors. E-mail: taimurahmad1@gmail.com, sharath.sriram@gmail.com (S.S.), omid.kaveh@gmail.com (O.K.)

## Content

## S1. Compositional and electronic analyses of STO$_x$ thin films

The chemical composition of the sputtered STO$_x$ thin film is analysed by X-ray photoelectron spectroscopy (XPS). In order to evaluate the stoichiometry, the core-level elemental spectra of sputtered STO$_x$ thin film are compared with the spectra collected from a reference stoichiometric STO (100) substrate, as shown in **Figure S1**a-c. The Sr 3$d$ and O 1$s$ spectra of both samples show similar binding energies which are within the margin of measurement error ($\pm$0.1 eV). The Sr 3$d$ spectra of stoichiometric STO and STO$_x$ (Figure S1a) are fitted into two split Sr 3$d_{5/2}$ orbital components corresponding to Sr$^{2+}$ binding energies at 132.94 eV and 132.97 eV, respectively.[1-2] The O 1$s$ spectra of stoichiometric STO and STO$_x$ (Figure S1b) are fitted with two distinct oxygen species. The peaks at binding energies of 529.45 eV (for stoichiometric STO) and 529.47 eV (for sputtered STO$_x$) correspond to O$^{2-}$ ions,[1] while peaks at higher binding energies are associated with the C–O bonds[1-3] arising from adventitious carbon on the surfaces. The Ti 2$p$ spectra of the both stoichiometric STO reference substrate and our STO$_x$ are shown in Figure S1c. The Ti 2$p$ spectrum of the stoichiometric STO suggests that Ti is present in its single oxidation state, *i.e.*, Ti$^{4+}$ with Ti 2$p_{3/2}$ peak at 458.37 eV.[2, 4-5] However, the resolved Ti 2$p$ spectrum of the sputtered STO$_x$ thin film show two distinct oxidation states with Ti 2$p_{3/2}$ peaks at binding energies of 457.9 eV and 456.2 eV. These correspond to the Ti$^{4+}$ and Ti$^{3+}$ oxidation states, respectively.[1, 3] The relative concentration of Ti$^{4+}$ and Ti$^{3+}$ species in STO$_x$ are calculated to be 70.6% and 29.4%, respectively, by integrating the respective fitted peaks. This indicates that the sputtered STO thin films are oxygen-deficient. This can be associated with the sputtering conditions of STO in a pure argon environment where Ar$^+$ ion bombardment results in the preferential removal of oxygen atoms and creates inherent oxygen vacancies.[6]

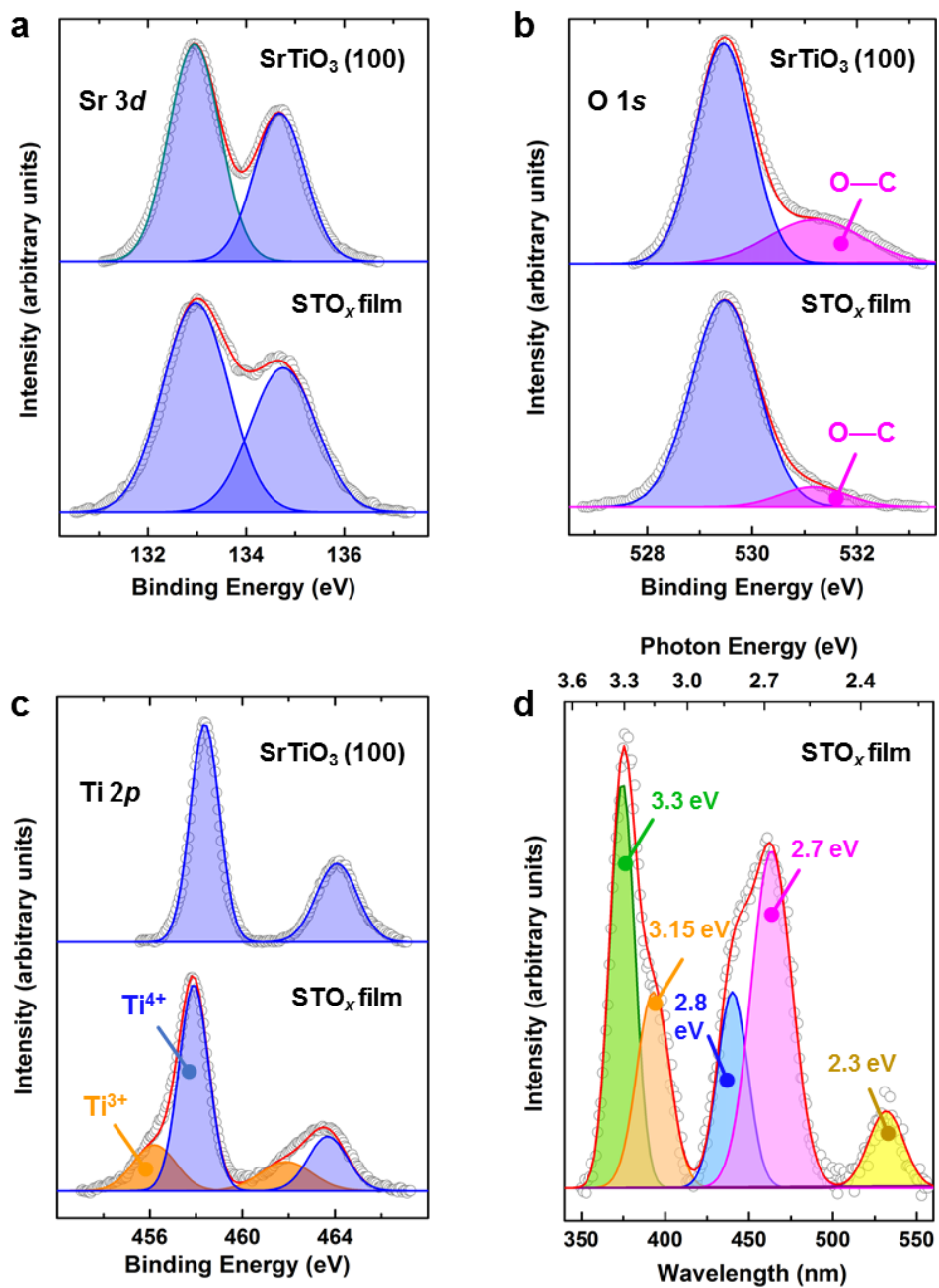

**Figure S1. Material characterisation of SrTiO₃.** The core-level resolved XPS spectra of (a) Sr 3*d* (b) O 1*s* and (c) Ti 2*p* collected from a reference stoichiometric SrTiO$_3$ (100) single crystal and the sputtered STO$_x$ thin film. (d) The resolved photoluminescence emission spectra of STO$_x$ collected at 325 nm excitation wavelength.

In order to assess the effect of as-grown oxygen vacancies on the electronic structure of $STO_x$, photoluminescence (PL) spectra are obtained using a 325 nm (3.82 eV) excitation source (Figure S1d). The PL emission from undoped STO (at room temperature) is associated with the presence of $TiO_5$ defects.[6-9] Furthermore, it is know that sputter deposition in a pure Ar environment creats as-grown oxygen vacancies (*i.e.*, $TiO_5$ defects),[9-11] which is also observed in our XPS analysis (explained above). As such, the PL emission is expected from $STO_x$ thin film. Figure S1d shows the PL spectra collected from bare $STO_x$ thin film at room temperature. The resolved PL spectra show a range of emission components between 2.3 eV and 3.3 eV which can be associated with the recombination of free electrons with in-gap levels and self-trapped exciton recombination.[6, 12] This indicates that as-grown oxygen vacancies introduce in-gap states which result in the overall bandgap reduction of $STO_x$. As such, the reduced bandgap of $STO_x$ may provide an assistive role in electroforming-free resistive switching behavior at relatively low applied bias than previously reported STO memristors.[13-14]

## S2. Electrical characterisation of STO$_x$ memristive devices

**Figure S2**a shows the microscopic photographs of the fabricated STO$_x$ MIM devices with different device area. Figure S2b shows resistances of the pristine STO$_x$ devices with respect to their active area. The resistances of at least 15 devices, with same active area are measured under a read voltage ($V_{READ}$) of 0.1 V.

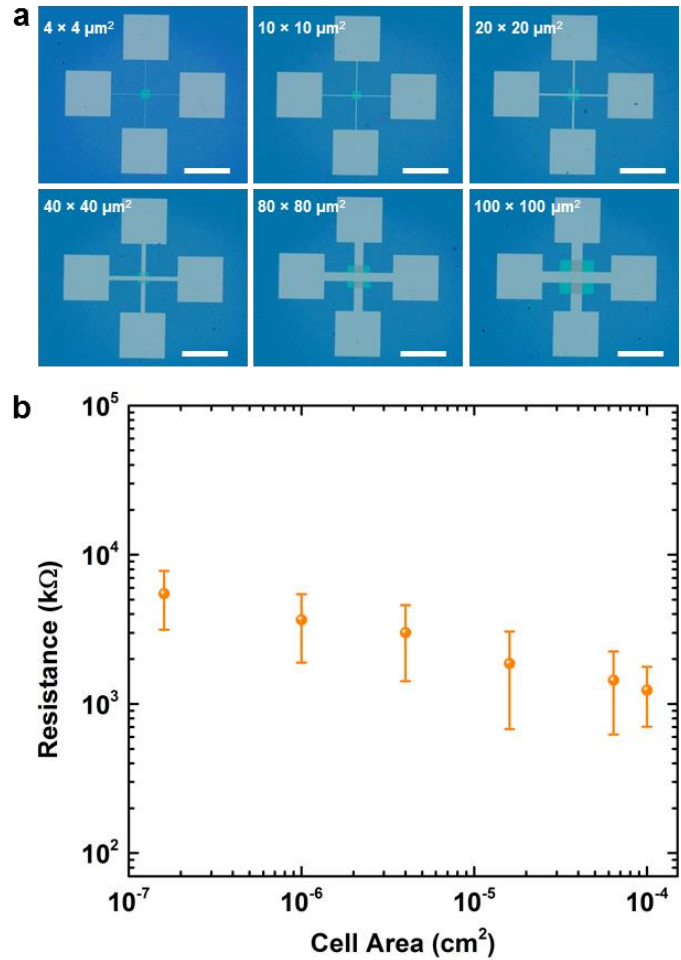

**Figure S2. Electrical characterization of the pristine STO$_x$ devices.** (a) Microscopic photographs of the fabricated STO$_x$ devices with different cell areas. Scale bar 400 µm. (b) The pristine resistances measured for at least 15 devices of the same cell area at $V_{READ}$ of 0.1 V. The error bars show the standard deviation in the measurements.

**Figure S3a** shows the representative *I–V* characteristics of first SET/REST curves of devices with different cell area. The statistical analysis (**Figure S3b**) of $V_{SET}$/$V_{RESET}$ for at least 15 MIM devices with similar cell area shows that the switching voltages decrease with increasing cell size. This can be associated with the decrease in pristine resistance with increasing cell size as shown in Figure S2b.

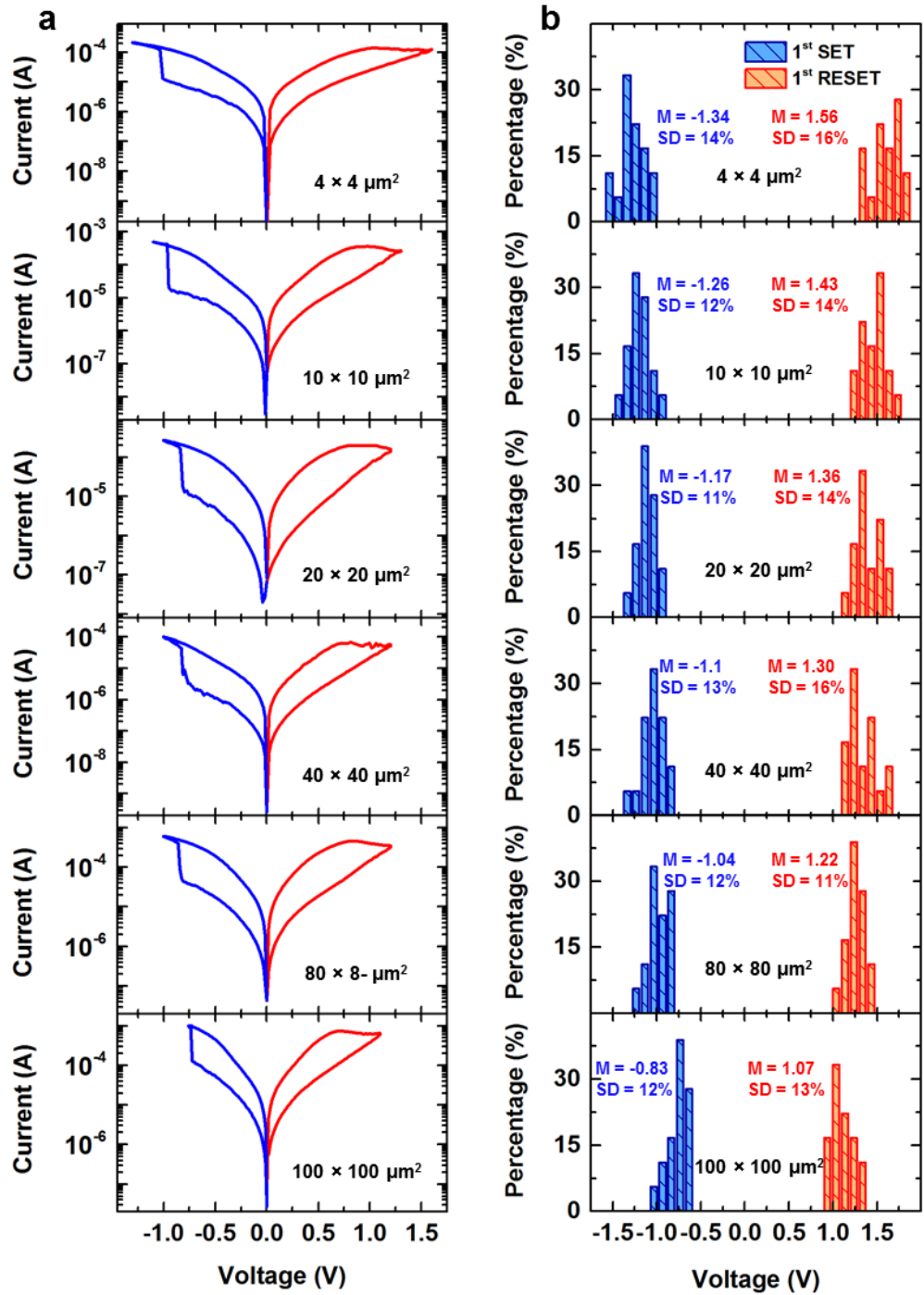

**Figure S3. First SET/RESET characterisation of the STO$_x$ devices.** (a) The representative $I$–$V$ characteristics of the first SET/RESET sweeps. The current compliance is set at ≥0.5 mA during the $I$–$V$ sweeps. (b) The statistics of the first $V_{SET}$ and $V_{RESET}$ of at least 15 MIM devices with same cell sizes (M: mean value and SD: standard deviation).

During the endurance evaluation of $STO_x$ MIM devices, train of WRITE/READ/ERASE/READ voltage pulses are used. **Figure S4** shows the characterisation of applied voltage pulses.

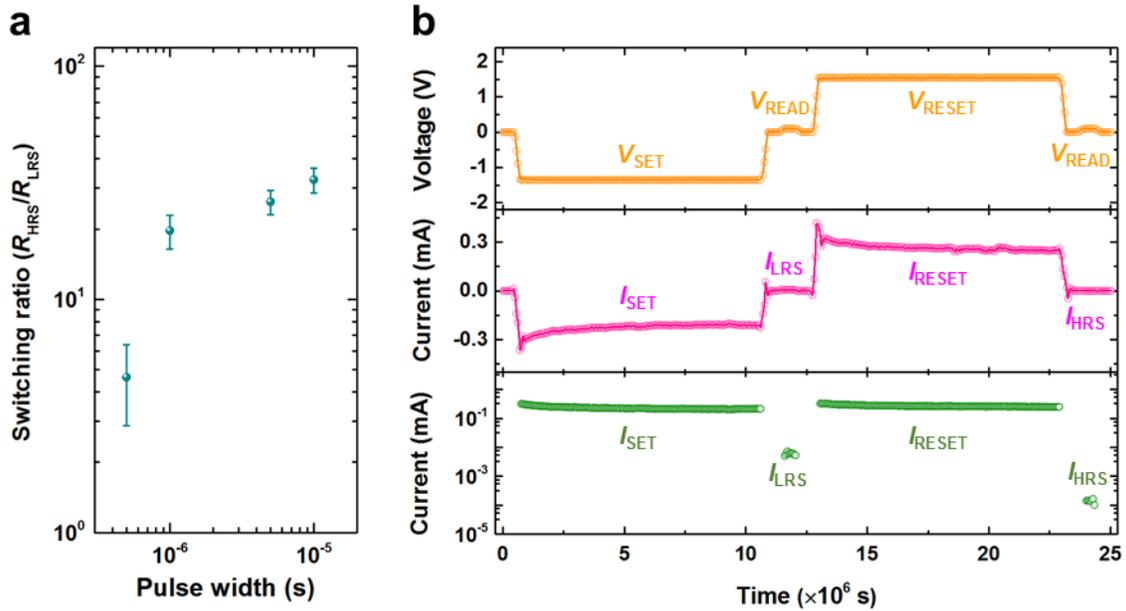

**Figure S4. Characterisation of SET/RESET pulses for the endurance of $STO_x$ MIM devices.** (a) Effect of SET and RESET pulse width on switching ratio. The READ pulses with amplitude of 0.1 V and fixed width of 200 ns are used to measure the switching ratios of 15 MIM devices with $10\times10$ µm² active area. (b) Response of the MIM device to the input $V_{SET}$ (-1.4 V)/$V_{RESET}$ (+1.6 V) voltage pulses with 1 µs pulse width (500 ns rise and fall time) and 0.1 V $V_{READ}$ pulses with width of 200 ns. The upper panel shows the input voltage pulses, middle panel shows measured current response to the applied voltage pulses on a linear scale while the lower panel shows only SET/RESET and LRS/HRS read currents on a log scales.

## S3. Cross-sectional analyses of $STO_x$ devices

Transmission electron microscope (TEM) and electron energy loss spectroscopy (EELS) techniques are used to analyze the morphology and composition of the $STO_x$ MIM devices. **Figure S5**a shows a TEM micrograph of the pristine $STO_x$ MIM device. During the TEM and live-FFT observation of the pristine devices, no noticeable crystalline regions are identified in the top Ti or the $STO_x$ functional oxide layer. To assess the electronic composition of the pristine device, the EELS area map and Ti–$L_{2,3}$ and O–$K$ edge profiles are obtained from a line-scan across the MIM structure (Figure S5b,c respectively). The EELS O–$K$ area map (Figure S5b) shows the presence of low oxygen content in the $STO_x$ oxide layer which indicates its oxygen deficient stoichiometry. The EELS Ti–$L_{2,3}$ edge profiles collected along a line scan (Figure S5c) show broad Ti–$L_3$ and Ti–$L_2$ peaks at the top Ti/$STO_x$ interface indicate the presence of mixed $Ti^{2+}$ and $Ti^{3+}$ oxidation states.[15] In the functional oxide layer, weak splitting of the $t_{2g}$ and $e_g$ peaks indicate $Ti^{4+}$ oxidation state. However, O–$K$ edge profiles are weak and noisy which makes difficult to clearly distinguish the fine structures and cannot be used to accurately identify the Ti valence.

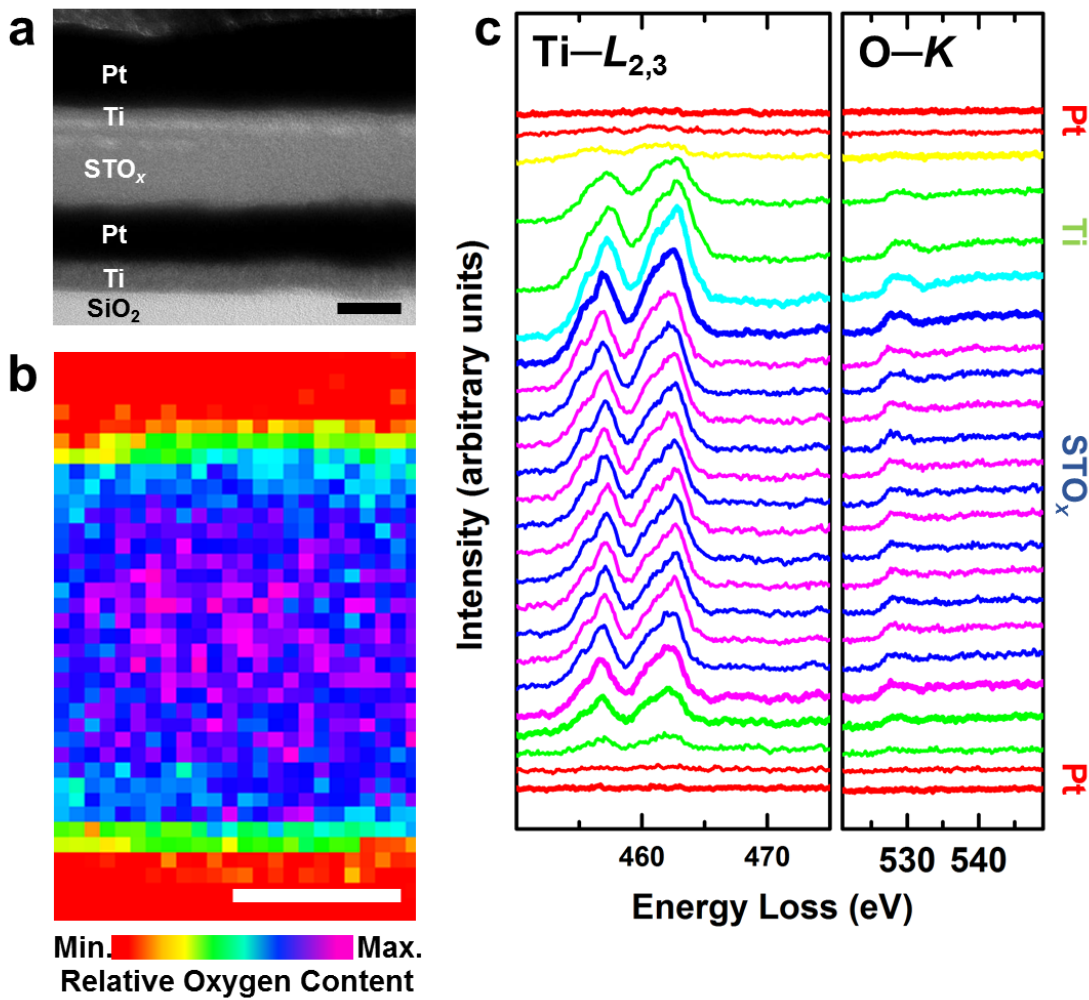

**Figure S5. Microstructure and electronic structure of the pristine STO$_x$ devices.** (a) TEM cross-section of a pristine device. Scale bar 20 nm. (b) The EELS O–$K$ edge area map of a pristine device. Scale bar 20 nm. (c) The EELS Ti–$L_{2,3}$ and O–$K$ edge profiles along a line scan across the pristine device.

## S4. Spike-time conversion to voltage

The time-to-digital-to-voltage circuitry converts spike-timing information into corresponding voltage magnitude. The corresponding voltages are applied to the bottom or top electrode of the $STO_x$ synaptic devices, depending on the sign of $\Delta t_1$ (potentiation or depression). **Table S1** lists the selected $\Delta t_1$ values and corresponding voltage amplitudes.

**Table S1. Conversion of the spike-timing information to voltage**. Selected values of $\Delta t_1$ are simulated to obtain corresponding voltage amplitudes.

| Depression | | Potentiation | | | |
|---|---|---|---|---|---|
| $\Delta t_1$ (ms) | $\Delta V$ (V) | $\Delta t_1$ (ms) | $\Delta V$ (V) | | |
| | | | $\Delta t_2 = 10$ ms | $\Delta t_2 = 80$ ms | $\Delta t_2 = 160$ ms |
| -45 | 0.500 | 3 | -0.800 | -0.900 | -1.200 |
| -42 | 0.529 | 6 | -0.778 | -0.875 | -1.167 |
| -39 | 0.557 | 9 | -0.757 | -0.850 | -1.135 |
| -36 | 0.586 | 12 | -0.735 | -0.825 | -1.103 |
| -33 | 0.614 | 15 | -0.714 | -0.800 | -1.071 |
| -30 | 0.643 | 18 | -0.692 | -0.775 | -1.039 |
| -27 | 0.671 | 21 | -0.671 | -0.750 | -1.007 |
| -24 | 0.700 | 24 | -0.650 | -0.725 | -0.975 |
| -21 | 0.729 | 27 | -0.628 | -0.700 | -0.942 |
| -18 | 0.757 | 30 | -0.607 | -0.675 | -0.910 |
| -15 | 0.786 | 33 | -0.585 | -0.650 | -0.878 |
| -12 | 0.814 | 36 | -0.564 | -0.625 | -0.846 |
| -9 | 0.843 | 39 | -0.542 | -0.600 | -0.814 |
| -6 | 0.871 | 42 | -0.521 | -0.575 | -0.782 |
| -3 | 0.900 | 45 | -0.500 | -0.550 | -0.750 |

**Table S2. Comparison of hippocampal data set and STO$_x$ memristors.** The synaptic weight change corresponding to different spike time differences is listed for both 2-pre-1-pre and 1-pre-2-post triplet pairing configurations. The hippocampal data set is taken from Ref.[16]

| Triplet pairing | Timing difference | | Weight change | |
|---|---|---|---|---|
| | $\Delta t_1$ | $\Delta t_2$ | Hippocampal experiments | STO$_x$ memristors |
| **2-pre-1-post** | 5 | -5 | -0.01±0.04 | -0.01±0.02 |
| | 10 | -10 | 0.03±0.04 | 0.01±0.02 |
| | 15 | -5 | 0.01±0.03 | 0.01±0.02 |
| | 5 | -15 | 0.24±0.06 | 0.28±0.03 |
| **1-pre-2-post** | -10 | 10 | 0.34±0.04 | 0.33±0.05 |
| | -5 | 5 | 0.33±0.04 | 0.33±0.05 |
| | -5 | 15 | 0.22±0.08 | 0.23±0.04 |
| | -15 | 5 | 0.29±0.05 | 0.31±0.06 |

**Table S3. Comparison of visual cortex data set and STO$_x$ memristors.** The synaptic weight change corresponding to different spike rates is listed for $\Delta t = 10$ ms and $\Delta t = -10$ ms. The visual cortex data is taken from Ref.[16]

| Spike rate ρ (Hz) | Cortex | | STO$_x$ memristors | |
|---|---|---|---|---|
| | $\Delta t = 10$ ms | $\Delta t = -10$ ms | $\Delta t = 10$ ms | $\Delta t = -10$ ms |
| 0.1 | -0.29±0.08 | -0.04±0.05 | -0.28±0.03 | -0.006±0.02 |
| 10 | -0.41±0.11 | 0.14±0.10 | -0.43±0.04 | 0.13±0.03 |
| 20 | -0.34±0.10 | 0.29±0.14 | -0.35±0.04 | 0.31±0.05 |
| 30 | 0 | 0.4±0 | 0±0.01 | 0.37±0.04 |
| 40 | 0.56±0.32 | 0.53±0.11 | 0.37±0.04 | 0.37±0.04 |
| 50 | 0.75±0.19 | 0.56±0.26 | 0.37±0.04 | 0.37±0.04 |

## S5. Event digitizer and time-to-voltage converter

The T2D (time-to-digital) module of our proposed CMOS circuit includes a timing control unit and a decoder (shown in **Figure S6**a). The timing control unit is a fully digital unit that receives pre and post digital spikes and generates timing interval signals solely based on counters. It can be configured for multiple protocol implementation and $\Delta t$ detection.[17] The resolution of timing detection for any $\Delta t$ is identified by (1) meaningful changes in synaptic weight ($\Delta w$) as the result of applying slight changes in voltage, and (2) resolution of overall time-to-voltage ($V_w$) conversion which is mainly depends on the DAC resolution. The T2D module passes a multi-bit spike timing (ST) digital signal to a decoder where two multi-bit horizontal/vertical select (HS/VS) digital signals are generated to adjust the number of resistors in series in our resistive DAC. The ST signal contains information about $\Delta t_1$ and $\Delta t_2$ and their different configurations which then are translated into an equivalent voltage ($V_w$) to be generated. It also includes flags that are part of $Sel_R$ and $Sel_C$ and identify whether a $\Delta t$ is positive or negative, hence, applying $V_w$ to the top or the bottom electrodes.

Figure S6b shows schematic of a single cell with two polysilicon resistors, controls and input/output signals. For a *k*-bit DAC, $2^k$ resistors are required. More detail on the implementation of the DAC and voltage follower/buffer (VF) is provided in Ref.[18-20]

Figure S6c illustrates generation of an internal reference voltage ($V_{ref}$) by using a voltage divider and also demonstrates a regulated current mirror to generate and regulate a fixed reference current to the chain of DAC resistors. The signal *ENA* provides the option to minimize the static current flowing through the resistor chain when the time-to-voltage circuit is disabled. It is worth highlighting again that these signals are affecting $Sel_R$ and $Sel_C$ to select a device in the array. As stated, voltage $V_{DAC}$ identifies maximum required voltage for programming the $STO_x$ memristor. In case of $V_{DAC}$ = 700 mV, 6-bit resistive DAC, and 1.2 V supply voltage in 90 nm standard CMOS technology, a 543 mV dynamic range on $V_w$, 191 µA active mode current, 99 nA standby mode leakage current at room temperature, and a 235 mV/µs slew rate is achieved.[18] The circuit also demonstrates a strong accuracy of ±5 mV with 8.5 mV step sizes. While we have modified the design, the original design is reported to have an area of around 175×175 µm$^2$ capable of driving up to 1 mm$^2$ of digital IP block.[18] We have added a fully digital timing control unit and removed $n$-well bias generation. While the main analog components are still part of the circuit, we estimate an area reduction of at least 20% is achievable in the modified time-to-voltage circuitry in comparison with the original body bias generator circuitry.

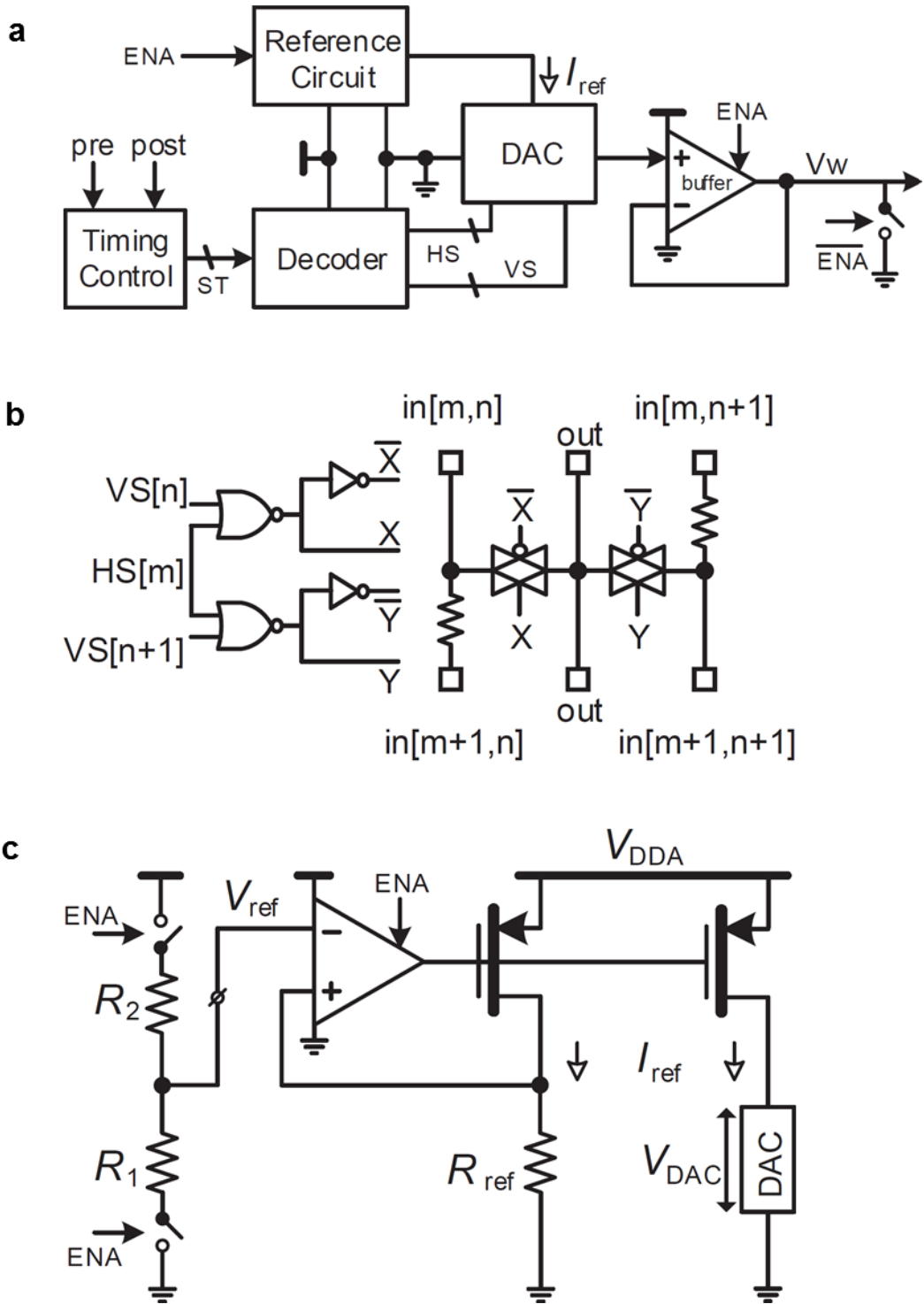

**Figure S6. Event digitization and Time-to-Voltage conversion.** Schematics of (a) time-to-digital module, (b) a single cell of resistive DAC and (c) a circuit to generate and regulate reference current and voltage for DAC. Circuit structures are those introduced in [19] for adaptive body biasing but internally modified to support current application to nano-arrays.